\documentclass[fleqn]{annalen}
\newcommand{\volume}{x}              
\newcommand{\xyear}{2002}            
\newcommand{\issue}{x}               
\newcommand{\recdate}{dd.mm.yyyy}    
\newcommand{\revdate}{dd.mm.yyyy}    
\newcommand{\revnum}{0}              
\newcommand{\accdate}{dd.mm.yyyy}    
\newcommand{\coeditor}{ue}           
\newcommand{\firstpage}{1}           
\newcommand{\lastpage}{x}            
\newcommand{\keywords}{acceleration, length measurement} 
\newcommand{\PACS}{03.30.+p, 04.20.Cv}
\newcommand{\shorttitle}{B.\ Mashhoon, et al., Length measurement in accelerated
systems}
\usepackage{amsmath, amssymb}
\usepackage{url}
\usepackage[dvips]{graphicx}
\usepackage[usenames, dvips]{pstcol}
\usepackage{pst-node, pst-plot}
\let\del=\partial

\let\theta=\vartheta
\newcommand{\nfrac}[2]{\leavevmode\kern.1em%
\raise.5ex\hbox{\scriptsize \ensuremath{#1}}%
\kern-.1em/\kern-.15em%
\lower.25ex\hbox{\scriptsize \ensuremath{#2}}}
\newenvironment{quote}
               {\list{}{\rightmargin\leftmargin}%
                \item\relax}
               {\endlist}

\begin{document}
\title{Length measurement in accelerated systems}

\author{Bahram Mashhoon and Uwe Muench} 

\newcommand{\address}{Department of Physics \& Astronomy, University
of Missouri--Columbia, Columbia, Missouri 65211, USA}
\newcommand{\email}{\tt MashhoonB@missouri.edu} 
\maketitle

\begin{abstract} 
We investigate the limitations of length
measurements by accelerated observers in Minkowski spacetime brought
about via the hypothesis 
of locality, namely, the assumption that an accelerated observer at
each instant is equivalent to an otherwise identical momentarily
comoving inertial observer. We find that consistency can be achieved
only in a rather limited neighborhood around the observer with linear
dimensions that are negligibly small compared to the characteristic
acceleration length of the observer. 
\hfill {\em Files length1.tex, 2002-06-24}
\end{abstract}

\section{Introduction}
The primary measurements in physics are the determinations of spatial
distances and temporal durations that are associated with the
effective establishment of a sufficiently local frame of
reference. This process involves macrophysical determinations
associated with the fact that physical observers and their frames of
reference obey the laws of classical (i.e.\ nonquantum) physics. The
basic nongravitational laws of physics refer to ideal inertial
observers; their measurements are briefly discussed in section
\ref{sec:op-length}. On the other hand, actual observers are all (more or
less) noninertial, i.e.\ accelerated. In fact, most experiments are
performed in laboratories fixed on the Earth, which---among other
motions---rotates about its axis; therefore, it is necessary to give a
theoretical description of the measurements of accelerated
observers. This is done via the hypothesis of locality described in
section \ref{sec:sec3}. This hypothesis in effect replaces the
accelerated observer by a continuous infinity of hypothetical
momentarily comoving inertial observers. Sections \ref{sec:sec4} and
\ref{sec:sec5} deal with the measurement of length by observers
undergoing translational and rotational accelerations,
respectively. Section \ref{sec:sec6} contains a discussion of our
results. 

\section{Simultaneity and length measurements}
We begin by reviewing some basic concepts and terms about length
measurement that are commonly used for inertial systems in special
relativity (SR).

An \emph{event} in SR is associated with a \emph{single} location in
space and a \emph{single} instant in time. The \emph{position of an
event} is defined to be the coordinate label on a rigid ruler that extends
from the spatial origin to the event; this notion is then naturally
extended to the spatial coordinates that characterize the location of
the event in space. The ruler is envisioned to extend indefinitely 
from some chosen origin. Such a choice is only possible in a global
inertial coordinate frame, which can be defined only in Minkowski
spacetime for inertial observers. The \emph{time of an event} is most
naturally defined as the reading on a clock located at the event's
position at the instant at which the event occurs. The rulers and
clocks used by an inertial observer are at rest relative to the
observer. Time is somehow a difficult notion to grasp, especially when
it becomes frame dependent under Lorentz transformations
\cite{rachel}.


\subsubsection*{Simultaneity}
All inertial observers in SR are assumed to be either actual
\emph{observers} or measuring devices that use synchronized clocks. To
determine the time of a 
distant event, an observer corrects for the travel time of a signal
originating at the event. To perform this correction the observer has
to know the distance to the event by either determining the event's
spatial coordinates in its reference frame or by prior measurement of the
distance. The determination of the location and the time of an
event are independent of the position of an observer compared to all
other observers in the same reference frame. 

The time ordering of the events depends on the relative
velocity of the inertial observers and the relative position of the
events, but not the positions of the observers since global
synchronization of clocks is assumed. The invariance of the
speed of light $c$ has an additional immediate implication: Two events
at different locations that occur at the same time in a given inertial
frame are not simultaneous in any other inertial frame. Moreover,
$v < c$ for any observer implies that the causal sequence of events is
independent of the inertial observers.  

\subsubsection*{Length measurements}
An inertial frame is globally defined, since the lifetime of clocks can be
ideally extended indefinitely and the rulers ideally extend indefinitely in
space. Hence, lengths are simply determined by the differences 
of the coordinate positions of the endpoint of line segments at the
same time in such a reference frame, i.e.\ $L=|\vec{x}_2-\vec{x}_1|$
is the length of the straight line segment extending from $\vec{x}_1$
to $\vec{x}_2$. In effect, the homogeneity and isotropy of spacetime in an
inertial frame allows us to sum intervals of time and space
corresponding to the use of finite clocks and rulers. 

A ruler of length $l_0$ at rest in an inertial frame contracts by a
factor of 
\begin{equation} \gamma^{-1} = \sqrt{1-\beta^2} \end{equation} 
as measured by standard observers at rest in an inertial frame moving
with speed $v=\beta c$ along the direction defined by the ruler; this effect
is known as the Lorentz-Fitzgerald contraction.


It is possible to define the distance between two inertial observers
using electromagnetic signals:\label{sec:op-length} One observer at
rest at $\vec{x}_1$ in some 
inertial frame sends out a light signal towards a second (possibly
moving) observer. The second observer at $\vec{x}_2$ sends a 
light signal back immediately after reception of the first light
signal. The first observer determines the time difference $\Delta t$ 
between sending the first light signal out and receiving the second
light signal at its position. The length between the observers is then
given by 
\begin{equation}
L^* := \frac{1}{2}\,c\,\Delta t\;.
\end{equation}
This length definition relies only upon the assumption that the speed
of light is constant and equal to $c$ in all inertial reference
frames; moreover it is 
consistent with the measurement of length based on rulers (i.e.\ $L^*=L$). 

\subsubsection*{Translational and rotational accelerations}
An inertial observer is an ideal that cannot be realized in
practice. All actual observers are accelerated. To develop the theory
of accelerated systems, let us define an orthonormal frame field
$e_\alpha$ 
for an accelerated observer. The components of the 
frame field are $\lambda^\mu{}_{(\alpha)}:=e^\mu{}_\alpha$, where
$e_\alpha = e^\mu{}_\alpha \del_\mu$. We choose $e_0$ to be the unit
vector $u^\mu(\tau) := \frac{1}{c} \frac{dx^\mu}{d\tau}$ that is
tangent to the worldline 
at a given event $x^\mu(\tau)$ and we parametrize the remaining frame
vectors characterizing the spatial directions also by $\tau$, which is
a temporal parameter measured along the accelerated path by \emph{the 
standard (static inertial) observers in the underlying global inertial
frame according to the formula} $\tau=\int \sqrt{1-\beta^2(t)}\,dt$.

The condition of orthonormality for the frame field reads
\begin{equation} \eta_{\mu\nu} \lambda^\mu{}_{(\alpha)}(\tau)\,
\lambda^\nu{}_{(\beta)}(\tau) \, = \eta_{\alpha\beta} =
\text{diag}(-1,+1,+1,+1) \;. \end{equation}

The derivative of the frame field along the accelerated path can be
expressed in the frame basis:
\begin{equation} \frac{d\lambda^\mu{}_{(\alpha)} }{d\tau} = \Phi_\alpha{}^\beta
(\tau) \lambda^\mu{}_{(\beta)}\;. \label{eq:4}\end{equation}
Using the orthonormality condition,
we find that 
$\Phi_{\alpha\beta}$ is antisymmetric
\begin{equation} \Phi_{\alpha\beta}(\tau) = - \Phi_{\beta\alpha}(\tau)
\;;\label{eq:eq5}\end{equation} 
we therefore define 
\begin{equation} \Phi_{\alpha\beta} := \left[\;\; \begin{matrix} 
0 \\
\hline
\\
- \vec{a}/c\\\\
\end{matrix}\; \right|\left.
\begin{matrix} 
  & \vec{a}/c &  \\\hline
 \\
 & \vec{\Omega} \\
\\
\end{matrix} \;\; \right] \label{eq:6}\;\;,\end{equation}
where $\Phi_{0i} = a_i/c$ and $\Phi_{ij} =
\epsilon_{ijk} \Omega^k$. Here $\vec{a}$ represents the ``electric''
component and is the translational acceleration, while $\vec{\Omega}$
represents the ``magnetic'' component and is the rotational
frequency of the local spatial frame (with respect to the local
nonrotating, i.e.\ Fermi-Walker transported, axes). 

\begin{figure}
\begin{center}\input{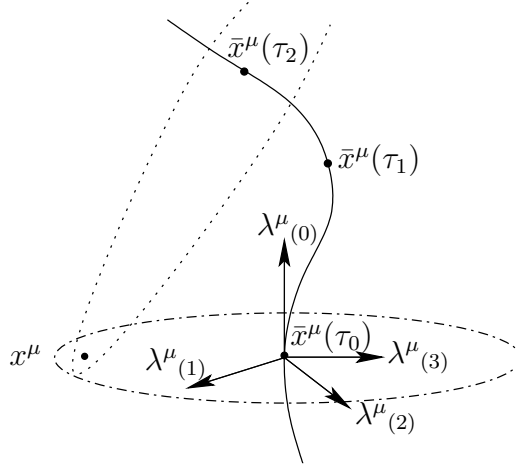}\end{center}
\caption{An event $x^\mu$ as seen by the observer
$\bar{x}^\mu(\tau_0)$ with its frame field
$\lambda^\mu{}_{(\alpha)}$. The geodesic coordinate system
$X^\mu=(c\tau,\vec{X})$ is limited
in space: If we
go beyond the time $\tau_1$, for example, coordinate assignments would
start to overlap, as shown for the time $\tau_2$. Since this cannot be
accepted, spatial coordinates have to be limited in general. Thus the
geodesic coordinate system is in general valid in a sufficiently
narrow worldtube along the timelike worldline of the observer.}
\label{fig:fig3}
\end{figure}

Let us now introduce a geodesic coordinate system $X^\mu$ in the
neighborhood of the accelerated path. 
At any time $\tau$ along the accelerated worldline (see
figure~\ref{fig:fig3}), the hypersurface orthogonal to the worldline
is Euclidean space and one can describe some event on this
hypersurface at $x^\mu$ to be at
$X^\mu$, where $x^\mu$ and $X^\mu$ are connected via
$X^0=c\tau$ and
\begin{equation} x^\mu = \bar{x}^\mu(\tau) +
X^i\lambda^\mu{}_{(i)}(\tau) \;, \label{eq:7} \end{equation}
where $\bar{x}^\mu$ represents the position of the
accelerated observer.

From \eqref{eq:7} we can derive (compare also with \cite{hehl90} and
references therein) the relation
\begin{align} dx^\mu & = \frac{1}{c}\frac{d\bar{x}^\mu}{d\tau} dX^0 + dX^i
\lambda^\mu{}_{(i)} + X^i d\lambda^\mu{}_{(i)} \notag \\
& = \lambda^\mu{}_{(0)} dX^0 + dX^i \lambda^\mu{}_{(i)} +
\frac{1}{c} X^i \, dX^0 \Big[ \Phi_i{}^0
\lambda^\mu{}_{(0)} + \Phi_i{}^j
\lambda^\mu{}_{(j)} \Big] \notag \\
& = \left[ \left(1 + \frac{\vec{a}\cdot \vec{X}}{c^2} \right)
\lambda^\mu{}_{(0)} + 
\frac{1}{c}\left(\vec{\Omega} \times \vec{X}\right)^i
\lambda^\mu{}_{(i)} \right]  dX^0 + 
\lambda^\mu{}_{(i)}\, dX^i \;,
\end{align} 
and hence the metric is
\begin{multline} \label{eq:accel-metr} ds^2 = \eta_{\mu\nu} \,dx^\mu \,dx^\nu \\ =
- \left[\left( 1 +
\frac{\vec{a}\cdot \vec{X}}{c^2} \right)^2 - \frac{1}{c^2} \left( \vec{\Omega}
\times \vec{X}\right)^2\right] (dX^0)^2 + \frac{2}{c} \left( \vec{\Omega}
\times \vec{X}\right) \cdot d\vec{X} dX^0 + \delta_{ij}\,
dX^i\,dX^j\;. \end{multline} 
Since we started from a global inertial frame in Minkowski spacetime,
the spatial part of the line element yields Euclidean space with its
origin occupied by the accelerated observer. 

This set of coordinates is limited. If we follow the above procedure
for two different times of the accelerated observer, our new
coordinates may not be unique, see figure~\ref{fig:fig3}. Since we
cannot accept two sets of coordinates in the same system for one event, we
have to require that the laboratory be sufficiently small. The charts
for our coordinates cannot be global for accelerated observers. In
fact, such geodesic coordinates are admissible as long as 
\begin{equation}
\left( 1 + \frac{\vec{a}\cdot \vec{X}}{c^2} \right)^2 > \frac{1}{c^2}
\left(\vec{\Omega}\times \vec{X}\right)^2 \;.\label{eq:eq10}
\end{equation}
Thus in the discussion of the admissibility of the geodesic
coordinates, two independent acceleration lengths must be considered:
the translational acceleration length \nfrac{c^2}{a} and the
rotational acceleration length \nfrac{c}{\Omega} that appear in
equation \eqref{eq:eq10}. 

The acceleration radii are connected with the domain of applicability
of the geodesic coordinate system around the reference accelerated
observer. It turns out that these acceleration lengths have another
independent and much more fundamental significance in terms of the
\emph{local} measurements of the accelerated observer following the
reference trajectory \cite{mash89,mash90}. This basic issue is
discussed in section \ref{sec:sec3}. 

It is important to remark here that one may use other (more
complicated) accelerated coordinate systems; however, these have their
attendant difficulties \cite{marzlin}. A discussion of these problems
is beyond the scope of this paper; therefore, we limit our
considerations here to geodesic coordinate systems. 

\subsubsection*{Length scales for accelerated observers}
The translational and rotational ``accelerations'' $a_i$ and $\Omega^k$
depend in general on both the velocity and the acceleration of the
observer. We therefore construct the scalar invariants of the
antisymmetric tensor $\Phi_{\alpha\beta}$, which are then independent
of the (coordinate-dependent) velocity:
\begin{align}
I=\frac{1}{2c^2} \Phi_{\alpha\beta} \Phi^{\alpha\beta} = -
\frac{a^2}{c^4} + \frac{\Omega^2}{c^2} \;, \notag \\
I^* = \frac{1}{4c^2} \Phi_{\alpha\beta}^* \Phi^{\alpha\beta} =
- \frac{\vec{a}}{c^2} \cdot \frac{\vec{\Omega}}{c} \;, \label{eq:invariants}
\end{align}
where $\Phi_{\alpha\beta}^*$ is the dual of $\Phi_{\alpha\beta}$, i.\
e.\ $\Phi_{\alpha\beta}^* = \epsilon_{\alpha\beta\gamma\delta}
\Phi^{\gamma\delta}$.
We define the finite lengths $|I|^{-\nfrac{1}{2}}$ and
$|I^*|^{-\nfrac{1}{2}}$ as the
\emph{proper acceleration lengths}.

Let us now see how long these lengths are in typical situations on the
earth. For the translational acceleration length on the earth's surface
we get ($a=9.8\nfrac{\text{m}}{\text{s}^2}, \Omega=0$)
\begin{equation} \frac{c^2}{a} 
= 9.46 \cdot 10^{15} \,\text{m} \approx  1\,\text{ly} 
\;, \end{equation} 
and for the rotational acceleration ($a=0, \Omega=\Omega_{\oplus}$)
the result is  
\begin{equation} \frac{c}{\Omega} = 
4.1253 \cdot 10^{12}\, \text{m} \approx 27.5 \,\text{AU} \;.
\end{equation}
Thus far we have discussed space-time measurements carried out by
inertial observers. We must now consider the results of measurements
carried out by an accelerated observer; moreover, it is important to
see how such measurements are affected by the presence of an
acceleration length $\mathcal{L}$.

\section{The Hypothesis of Locality}
In a spacetime diagram an inertial observer can be portrayed as a
straight line. An observer that is linearly accelerated at some time
will have a curved worldline. What will this accelerated observer measure?
Typically, the \emph{Hypothesis of Locality} \cite{mash89,mash90} is
tacitly\label{sec:sec3} assumed: 
\begin{quote}
An accelerated observer measures the same physical results as a standard
inertial observer that has the same position and velocity at the time of
measurement. 
\end{quote}
The curved path of the observer is substituted by the straight line
tangential to the curve at the time of measurement. The radius of
curvature of the accelerated worldline is characterized by the
acceleration length $\mathcal{L}$; the hypothesis of  locality
therefore assumes that locally $\mathcal{L}=\infty$. It is necessary
to investigate if it is all right to reduce all measurements to
the linear approximation, especially if we leave the infinitesimal
neighborhood of an event and considering that realistic measuring
devices are not infinitesimal.

The hypothesis of locality originates from Newtonian mechanics of
classical point particles. The state of such a particle is given at
each instant of time by its position and velocity. It follows that the
hypothesis of locality is evidently valid in Newtonian mechanics and
this explains the fact that no new physical assumption is needed in
Newtonian physics to deal with accelerated systems. 

It is important to recognize that the hypothesis of locality is
crucial for the physical implementation of Einstein's heuristic
principle of equivalence. This cornerstone of general relativity and
the hypothesis of locality \emph{together} imply that an observer in a
gravitational field is pointwise inertial. 

A restricted hypothesis of locality is the so-called \emph{clock
hypothesis}, which is a hypothesis of locality only concerned about
the measurement of time. This hypothesis implies that a
\emph{standard} clock in fact measures $\tau$,
$d\tau=\sqrt{1-\beta^2(t)}\,dt$, along its path; $\tau$ is
then the \emph{proper} time along this accelerated path.
In the following sections, we set $\tau=0$ when $t=0$. 

According to most experiments, the hypothesis of locality seems to be
true. No experiment has yet shown the hypothesis of locality to be
violated (outside of radiation effects). The main reason for this
finding is that all relevant length scales in feasible experiments are
very small in relation to the huge acceleration lengths of the tiny
accelerations we usually experience. For instance, if we take the
wavelength of light for a typical laboratory optics experiment, $\lambda
\sim 10^{-7}$\,m, the factor $\nfrac{\lambda}{\mathcal{L}}$ is
around $10^{-23}$ and $10^{-20}$ for translational and rotational
accelerations, respectively. As long as all length scales are very small
compared to the acceleration lengths, it seems reasonable to assume
that differences between observations by accelerated and comoving inertial
observers will also be very small. 

It is the purpose of this paper to examine critically certain basic
aspects of the hypothesis of locality in connection with the
measurements of accelerated observers. To this end, we study in this
work the measurement of \emph{length} by accelerated observers. This
choice is based in two considerations: (1) length measurement is a
subject of crucial significance for a geometric theory of spacetime
structure and (2) the hypothesis of locality must be applied not just
at one event but at a continuous infinity of events for the
determination of a finite length. 

For practical purposes, the hypothesis of locality replaces the
accelerated observer by an infinite sequence of otherwise identical
momentarily comoving inertial observers. Every inertial observer is
endowed with a natural orthonormal tetrad frame in Minkowski
spacetime. Therefore, the same holds for an accelerated observer by
the hypothesis of locality. It is then natural to interpret the
results of section \ref{sec:op-length} as follows: The accelerated observer
carries an orthonormal frame $\lambda^\mu{}_{(\alpha)}(\tau)$ along
its trajectory such that at each instant of its proper time $\tau$,
the accelerated observer's temporal axis is $\lambda^\mu{}_{(0)}$ and
the spatial axes $\lambda^\mu{}_{(i)}$, $i=1,2,3$, characterize the
3-dimensional Euclidean space of this observer. Thus in the geodesic
coordinate system adapted to this tetrad frame, the spatial part of
the flat spacetime metric is always the 3-dimensional Euclidean space
as in equation \eqref{eq:accel-metr}. Moreover, the local acceleration
scales associated with the measurements of the observer are defined
via equations \eqref{eq:4}--\eqref{eq:6} and
\eqref{eq:invariants}. These have a \emph{physical} significance that
is distinct from the acceleration radii that mark the limits of the
validity of the accelerated coordinate system as can be made clear by
a simple example: For observers fixed on the rotating Earth,
Earth-based coordinates are essentially valid only up to the light cylinder
parallel to the Earth's axis and at a radius of
$\nfrac{c}{\Omega_{\oplus}} \approx 28\,\text{AU}$ from it. This light
cylinder, however, has no influence on the local measurements of the
observer and the reception of astronomical data on the Earth. In
contrast, the fact that such an observer is noninertial and therefore
has local acceleration scales associated with it does affect its
measurements as demonstrated by the phenomenon of spin-rotation
coupling \cite{mash95}. 

In our description of accelerated observers, an observer following a
straight worldline in an inertial frame is not necessarily
inertial. Consider, for example, an accelerated observer at rest in
Minkowski spacetime that refers its observations to rotating axes. The
observer's worldline is simply parallel to the time axis and the
limitation of a geodesic coordinate system established around this
observer does not arise from what is depicted in figure
\ref{fig:fig3}, but stems from the fact that observers at rest in the
rotating frame would be moving relative to the reference observer at
less than the speed of light only within its light cylinder. It
follows that in the treatment of accelerated (i.e.\ noninertial)
observers, the worldline as well as the spatial frame along the
worldline must be taken into account. A more satisfactory frame bundle
approach is indeed possible \cite{hehl90,Uwe00}, but such a treatment
is beyond the scope of the present paper.

In the following sections, we consider specific thought experiments
involving the measurement of distance between two accelerated
observers. 

\section{Linear acceleration}
\begin{figure}
\begin{center}\input{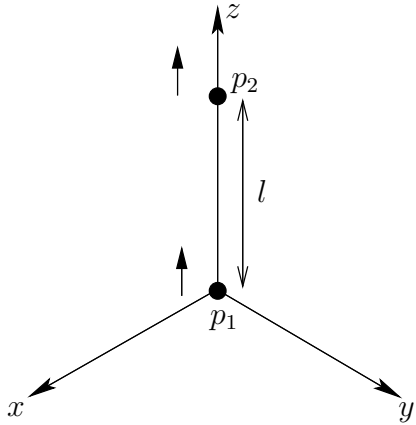}\end{center}
\caption{Two observers a distance $l$ apart start accelerating from
rest with identical acceleration profiles along the $z$-axis.}
\label{fig:fig4}
\end{figure}

Consider two observers that are at rest in an inertial frame and a
distance $l$ apart, see 
\cite{mash89,mash97a}. At $t=0$ they both start to accelerate the \emph{same}
way, according to a preplanned acceleration profile. This type of
thought experiment has been considered before \cite{bell}. We put one of the
objects at the origin of our inertial coordinate system and the other
one at $(0,0,l)$, and we assume that they accelerate linearly along
the $z$-direction. For later 
calculations, we will specify the acceleration to be uniform along the
$z$-axis, see figure~\ref{fig:fig4}. To avoid unphysical situations,
we assume that the acceleration is always turned off at some finite
time $t>0$.\label{sec:sec4} 

An inertial observer at rest in the inertial frame describes the
positions of the two accelerating objects to be 
\begin{equation}
z_{p_1}(t) = \int\limits_0^t v(t) \, dt \;,\quad z_{p_2}(t) = l +
\int\limits_0^t v(t) \, dt \;. \label{eq:14}
\end{equation}
Hence, the distance between the accelerating objects stays constant,
since $z_{p_2}(t) - z_{p_1}(t) = l$. 

Let us now investigate what comoving observers would measure for the
distance between $p_1$ and $p_2$. The hypothesis
of locality implies that both of the accelerated observers pass
through the same infinite sequence of momentarily comoving inertial
systems. The Lorentz transformation between the original inertial
system and one of the comoving systems gives
\begin{equation} l' = \frac{1}{\sqrt{1-\frac{v^2}{c^2}}} \;l = \gamma l \;, \end{equation}
which we generalize to
\begin{equation}
l' = \frac{1}{\sqrt{1-\frac{v^2(t)}{c^2}}} \; l = \gamma(t) l \;.\label{eq:l-inertial}
\end{equation}
This has a simple physical interpretation: The Lorentz-Fitzgerald
contracted distance between our accelerated objects is always $l$,
hence the actual distance between them must be larger by the momentary
Lorentz $\gamma$-factor. It is important to recognize that $p_1$ and $p_2$
could be any two points in a measuring device that is accelerated.

Specifically, let us imagine a set of accelerated observers populating
the distance between $p_1$ and $p_2$ undergoing exactly the same
motions as $p_1$ and $p_2$. At any given time $\hat{t}$, each of these
observers is pointwise equivalent to a comoving inertial observer in
accordance with the hypothesis of locality. The Lorentz transformation
connecting the global background inertial frame with the rest frame of
a comoving inertial observer at $(0,0,\hat{z})$ is given by
\begin{align} 
c(t-\hat{t}) & = \hat{\gamma} ( ct'+\hat{\beta} z')\;, \label{eq:l1}\\
x= x'\,,& \quad  y=y'\,,\quad 
z-\hat{z} = \hat{\gamma} (z'+ c \hat{\beta} \label{eq:l2}
t') \;, 
\end{align}
where $\hat{\beta}$ and $\hat{\gamma}$ refer to the common speed of
the system at $\hat{t}$. The consideration of length measurements of
the standard observers in their inertial frames then leads to equation
\eqref{eq:l-inertial}, i.\,e.\ the events $p_1:(c\hat{t},0,0,\hat{z_1})$ and
$p_2:(c\hat{t},0,0,\hat{z_2})$ in the background global frame correspond to
$p_1:(ct_1', 0,0,z_1')$ and $p_2:(ct_2',0,0,z_2')$, where
$ct_1'=\hat{\gamma}\hat{\beta}(\hat{z}-\hat{z_1})$,
$z_1'=-\hat{\gamma}(\hat{z}-\hat{z_1})$,
$ct_2'=\hat{\gamma}\hat{\beta}(\hat{z}-\hat{z_2})$, and
$z_2'=-\hat{\gamma}(\hat{z}-\hat{z_2})$; therefore,
$z_2'-z_1'=l'=\hat{\gamma} (\hat{z_2}-\hat{z_1})= \hat{\gamma}
l$.

For an alternative description, we should be able to replace the
infinite sequence of inertial systems by one system in a
continuously moving frame; for example, a coordinate system that has at
its spatial origin one of the accelerating objects ($p_1$). To this
end, it is useful to introduce at this point the simplifying assumption that the
observers are subject to \emph{uniform} acceleration $g$. Observer
$p_1$ thus follows a hyperbolic spacetime trajectory given by 
\begin{equation}
t=\frac{c}{g} \sinh \left(\frac{g\tau}{c}\right)\;,\quad
x=y=0\;,\quad
z=z_0+\frac{c^2}{g}\left(-1+\cosh\left(\frac{g\tau}{c}\right)\right)\;,\label{eq:19}
\end{equation}
where $z_0=0$ and $\tau$ is the proper time along the trajectory such
that $\tau=0$ at $t=0$. The speed of the observer is thus
$v=c\tanh\left(\frac{g\tau}{c}\right)$. We can construct an orthonormal tetrad
frame along the reference trajectory such that at each instant it would coincide
with the frame field of the momentary Lorentz transformation
\eqref{eq:l1} and \eqref{eq:l2},
\begin{subequations}
\begin{align}
\lambda^\mu{}_{(0)} &=
\left(\gamma,0,0,\gamma\beta\right)
\;,\\
\lambda^\mu{}_{(1)} &=
(0,1,0,0)
\;,\\
\lambda^\mu{}_{(2)} &=
(0,0,1,0)
\;,\\
\lambda^\mu{}_{(3)} &=
\left(\gamma\beta,0,0,\gamma\right)
\;,
\end{align}
\end{subequations}
where $\beta=\tanh\left(\frac{g\tau}{c}\right)$ and
$\gamma=\cosh\left(\frac{g\tau}{c}\right)$. It follows from the
hypothesis of locality that this is in fact the tetrad frame of the
accelerated observer. Using this tetrad frame in equations 
\eqref{eq:4}--\eqref{eq:6} reveals that $\vec{a}=(0,0,g)$ and
$\vec{\Omega}=\vec{0}$, so that the only proper acceleration length
associated with the observer is $\mathcal{L}=\frac{c^2}{g}$, as expected. The
spatial frame is in fact nonrotating, i.\,e.\ it is Fermi-Walker
transported along the trajectory, so that the geodesic coordinate
system constructed on this basis is a Fermi system. 

According to equation \eqref{eq:7}, the relationship between the
global inertial coordinates $x^\mu=(ct,x,y,z)$ and Fermi coordinates
$X^\mu=(cT,X,Y,Z)$ along $p_1$ is given by
\begin{equation} 
c t = \left( Z + \frac{c^2}{g}\right) \sinh \left( \frac{gT}{c}\right),\quad
x=X,\quad y=Y, \quad z= \left( Z + \frac{c^2}{g}\right) \cosh \left(
\frac{gT}{c}\right) - \frac{c^2}{g}+z_0\;,\label{eq:21}
\end{equation}
so that $p_1$ is always at the spatial origin of the Fermi system with
$T=\tau$ and $z_0=0$. 
If the positions of the two accelerating objects in the original
inertial frame at a time $\bar{t}$ are given by
$p_1:(c\bar{t},0,0,\bar{z})$ and $p_2:(c\bar{t},0,0,l+\bar{z})$, then
the corresponding positions in the moving coordinate system are
$p_1:(cT,0,0,0)$ and $p_2:(cT_2,0,0,L)$. From equation \eqref{eq:21} we get the
relations
\begin{equation}\label{eq:trans1}
c\bar{t} = \frac{c^2}{g} \sinh \left(\frac{gT}{c}\right),
\quad \bar{z} = \frac{c^2}{g} \left[
\cosh\left(\frac{gT}{c}\right)-1 \right] \end{equation}
and 
\begin{equation} c\bar{t} = \left(L+\frac{c^2}{g}\right) \sinh
\left(\frac{gT_2}{c}\right) ,\quad
\bar{z}+l=\left(L+\frac{c^2}{g}\right) \cosh
\left(\frac{gT_2}{c}\right) -\frac{c^2}{g} \;. \end{equation}
Using $\cosh^2\Theta-\sinh^2\Theta=1$ in the last equation yields
\begin{equation} \left(L+\frac{c^2}{g} \right)^2 = \left( l + \frac{c^2}{g} +
\bar{z} \right)^2 - c^2 \bar{t}^2 \;; \end{equation}
then, substituting for $\bar{t}$ and $\bar{z}+\frac{c^2}{g}$ using
\eqref{eq:trans1} leads to  
\begin{equation} \left(L+\frac{c^2}{g} \right)^2 =  l^2 + 2
l \frac{c^2}{g}\cosh\left(\frac{gT}{c}\right)   +
\left(\frac{c^2}{g}\right)^2 \;,\end{equation}
and this gives after some algebra 
\begin{equation} L = \frac{c^2}{g} \left[ \sqrt{1+2\epsilon\gamma+\epsilon^2} - 1
\right] = \frac{l'}{\gamma\epsilon} \left[
\sqrt{1+2\epsilon\gamma+\epsilon^2} - 1 \right]\label{eq:eq26} \end{equation}
with $\epsilon=\nfrac{l}{\frac{c^2}{g}}$ and
$\gamma=\cosh\left(\frac{gT}{c}\right)$. The parameter $\epsilon$
compares the length $l$ with the acceleration length in this case.
For $\epsilon \gtrsim 1$, equation  \eqref{eq:eq26} implies that $L$
and $l'$ can be very different; therefore, let us assume that
$\epsilon\ll 1$. 
We now can compare $L$ with $l'$, after applying the approximation
$\sqrt{1+x} = 1 +\frac{1}{2}x - \frac{1}{8}x^2 + \frac{1}{16}x^3 +
\mathcal{O}(x^4)$ for $|x|<1$,
\begin{equation} \frac{L}{l'} = 1 - \frac{1}{2} \beta^2\gamma\epsilon +
\frac{1}{2}\beta^2\gamma^2\epsilon^2 + \mathcal{O}(\epsilon^3)
\label{eq:29} \;. \end{equation} 
The length $L$ measured from $p_1$ in this accelerated frame
differs from the length $l'$ measured in a comoving inertial frame, if the
length $l$ is not negligibly small in comparison to the acceleration length. 

We now can change positions in this accelerated frame and investigate
what length is measured from position $p_2$. Observer $p_2$ also
follows a hyperbolic trajectory given by equation \eqref{eq:19} with
$z_0=l$. The corresponding transformation between inertial coordinates
and Fermi coordinates is given by \eqref{eq:21} with $z_0=l$. 
If the positions of the two accelerating objects in the original
inertial frame at a time $\bar{t}$ are now given as before by
$p_1:(c\bar{t},0,0,\bar{z})$ and $p_2:(c\bar{t},0,0,l+\bar{z})$, then
the corresponding positions in the moving Fermi coordinate system are
$p_1:(cT_1,0,0,-L')$ and $p_2:(cT,0,0,0)$. From equation \eqref{eq:21} we get the
relations
\begin{equation}\label{eq:trans2-1}
c\bar{t} = \left(\frac{c^2}{g}-L'\right) \sinh \left(\frac{gT_1}{c}\right),
\quad \bar{z}-l = \left(\frac{c^2}{g}-L'\right)
\cosh\left(\frac{gT_1}{c}\right)- \frac{c^2}{g} \end{equation}
and just as in equation \eqref{eq:trans1},
\begin{equation} \label{eq:trans2-2}
c\bar{t} = \frac{c^2}{g} \sinh
\left(\frac{gT}{c}\right) ,\quad
\bar{z}=\frac{c^2}{g} \cosh
\left(\frac{gT}{c}\right) -\frac{c^2}{g} \;. \end{equation}
Using $\cosh^2\Theta-\sinh^2\Theta=1$ in equation \eqref{eq:trans2-1} yields
\begin{equation} \left(\frac{c^2}{g} -L'\right)^2 = \left( \frac{c^2}{g} +
\bar{z} -l \right)^2 - c^2 \bar{t}^2 \;, \end{equation}
which after substituting for $\bar{t}$ and $\bar{z}+\frac{c^2}{g}$ using 
\eqref{eq:trans2-2} leads to  
\begin{equation} \left(\frac{c^2}{g}-L' \right)^2 =  l^2 - 2
l \frac{c^2}{g}\cosh\left(\frac{gT}{c}\right)   +
\left(\frac{c^2}{g}\right)^2 \;,\end{equation}
and this gives after some algebra 
\begin{equation} L' = \frac{c^2}{g} \left[ 1 - \sqrt{1-2\epsilon\gamma+\epsilon^2} \right] = \frac{l'}{\gamma\epsilon} \left[
1 - \sqrt{1-2\epsilon\gamma+\epsilon^2} \right] \end{equation}
with $\epsilon=\nfrac{l}{\frac{c^2}{g}}$ and 
$\gamma=\cosh\left(\frac{gT}{c}\right)$ as above. 
Again, for $\epsilon \ll 1$ let us now compare $L'$ with $l'$, after
applying the approximation 
$\sqrt{1-x} = 1 -\frac{1}{2}x - \frac{1}{8}x^2 - \frac{1}{16}x^3 +
\mathcal{O}(x^4)$ for $|x|<1$,
\begin{equation} \frac{L'}{l'} = 1 + \frac{1}{2} \beta^2\gamma\epsilon +
\frac{1}{2}\beta^2\gamma^2\epsilon^2 + \mathcal{O}(\epsilon^3)
\;. \label{eq:37} \end{equation}
The length $L'$ measured from $p_2$ in this accelerated frame
differs from the length $L$ (in fact, $L'$ is larger than $L$ 
for $0<\epsilon<1$) and from the length $l'$, if the length $l$
is not negligible compared to the acceleration length.   

Let us now take another approach, based on our operational definition of
length using electromagnetic signals: We want to measure the length by
timing light 
rays. The relation between the measured time and the length can then
be derived from the metric \eqref{eq:accel-metr} for our case:
\begin{equation}
 ds^2 = - \left( 1 + \frac{gX^3}{c^2} \right) ^2 (dX^0)^2 + \delta_{ij}
dX^i dX^j \;. \end{equation}
For light rays along the $X^3$- or $Z$-axis, $ds^2=0$, $dX^1=0$, and
$dX^2=0$, and therefore:
\begin{equation}
 dZ = \pm \left( 1 + \frac{gZ}{c^2} \right) c\,dT \;. \end{equation}
After integration we get
\begin{equation}
 c T + \text{constant} = \pm \frac{c^2}{g} \ln
\left(1+\frac{gZ}{c^2}\right) \;. \end{equation}
From the viewpoint of observer $p_1$, i.e.\ in the Fermi frame in
which $p_1$ is at rest, let us suppose that the signal is emitted at
time $T_1^-$ from $Z=0$ such that the light travels the distance $Z: 0
\to L$ and arrives at time $T_2$ at 
$p_2$, since that is the position of $p_2:(T_2,0,0,L)$ when the light
arrives, i.e.\ $c\ln(1+\nfrac{gL}{c^2})=g(T_2-T_1^-)$, and 
then back along $Z:L \to 0$, if we assume that the light is reflected by
$p_2$ without delay so that it returns to $p_1$ at $T_1^+$ such that
$c\ln(1+\nfrac{gL}{c^2})=g(T_1^+-T_2)$. Let us note that
$T_2=(T_1^++T_1^-)/2$, which is the standard synchronization
condition for distant events. With
$L^*=c (T_1^+-T_1^-)/2=\nfrac{c^2}{g}\ln(1+\nfrac{gL}{c^2})$
for the length determined by light-ray 
timing, we find that $L^*<L$, where $L$ is determined by rulers in the
accelerated system based on the hypothesis of locality; specifically,
we get using \eqref{eq:29} 
\begin{equation}  L^* = \frac{l'}{\gamma\epsilon}\ln \left( 1 +
\gamma\epsilon - 
\frac{1}{2} \gamma^2\beta^2\epsilon^2 + \mathcal{O}(\epsilon^3)
\right)\;. \end{equation}
With $\ln(1+x) = x-\frac{1}{2} x^2+\mathcal{O}(x^3)$ for $-1<x\le1$, we
finally find
\begin{equation}
 \frac{L^*}{l'} = 1 - \frac{1}{2} \gamma\epsilon(1+\beta^2) +
\mathcal{O}(\epsilon^2) \;,\end{equation}
yet another result for the measured length if $\epsilon \not\approx
0$.  

From the viewpoint of observer $p_2$, i.e.\ in the Fermi frame in
which $p_2$ is at rest, the thought experiment can be repeated by
sending the light signal from $p_2$ to $p_1$ 
and back without delay; in this case, a similar analysis holds except
that we have to use $L'$ instead of $L$ in the expression
corresponding to $L^*$. The calculation for this case yields using
\eqref{eq:37} 
\begin{equation} \frac{L'{}^*}{l'} = 1 - \frac{1}{2}
\frac{\epsilon}{\gamma} + \mathcal{O}(\epsilon^2) 
\;.\end{equation} 

It follows from these results that consistency can be achieved only if
$\epsilon=\nfrac{gl}{c^2} \ll 1 $ is below the level of sensitivity of
the measurements of the accelerated observers. 

It is possible to generalize our approach to arbitrary accelerated
systems: Imagine two observers that are initially at rest in an
inertial frame and subsequently move in exactly the same way for
$t>0$. A vector analogue of equation \eqref{eq:14} then implies that
$\vec{x}_{p_2}(t)-\vec{x}_{p_1}(t)=\vec{x}_{p_2}(0)-\vec{x}_{p_1}(0)$,
so that the Euclidean length between them remains the same as measured
in the inertial frame. The determination of the distance between them
as measured by the accelerated observers can be discussed as in the
foregoing treatment. On the other hand, it is more interesting to
consider a situation where the distance between the accelerated
observers is defined along a curve rather than a straight line such as
for two points fixed on the rotating Earth. Therefore, in the
following section we consider rotating observers and assume that the
rate of rotation is uniform for the sake of simplicity.

\section{Rotational acceleration}
\begin{figure}
\begin{center}\input{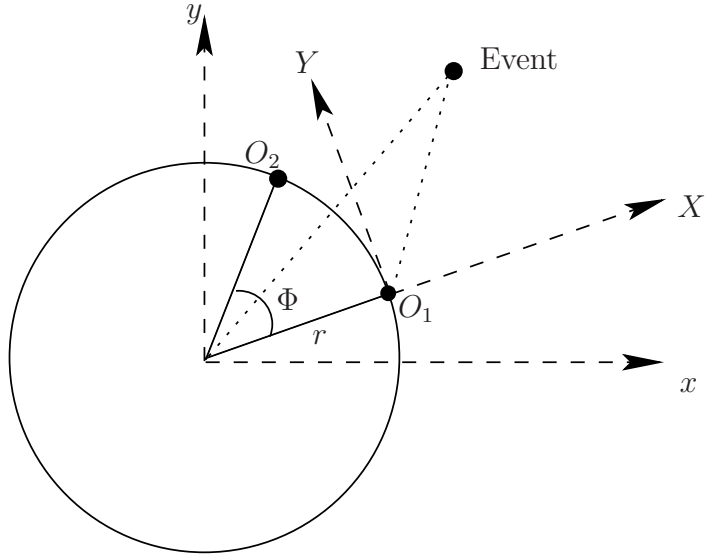}\end{center}
\caption{Two observers uniformly rotating on a circle of radius $r$
with azimuthal angles $\phi_1=\Omega_0t$ and $\phi_2=\Omega_0t+\Phi$. An 
event can be described in the inertial frame $(ct,x,y,z)$
and in a rotating geodesic coordinate system $(cT,X,Y,Z)$.}
\label{fig:fig5}
\end{figure}

We consider two observers $O_1$ and $O_2$ that rotate uniformly with angular
velocity $\Omega_0$ on a circle with radius $r$ and with a constant
angle $\Phi$ between them as in figure~\ref{fig:fig5}. 
An inertial observer at rest in the global inertial frame would
describe the arclength between 
the observers to have a constant length of $l=r\Phi$.\label{sec:sec5} 

Let us now again investigate what comoving observers measure.  For the
sake of concreteness, we imagine a set of rotating observers
populating the circle between $O_1$ and $O_2$ undergoing exactly the
same motions as $O_1$ and $O_2$. The
hypothesis of locality allows us to construct an infinite sequence of
momentarily comoving inertial observers tangential to particles on the
arc between the two circling observers. The Lorentz transformation
between the original inertial observers at rest and one of the comoving
inertial observers gives infinitesimally 
\begin{equation} dl' = \frac{1}{\sqrt{1-\frac{v^2}{c^2}}} \;dl = \gamma\, dl \;, \end{equation}
with $v=r\Omega_0$. While $\gamma$ in the case of uniform linear
acceleration was changing, it is constant here. By 
integrating over the comoving inertial observers we get $l'=\gamma l$
for the arclength between the objects. The physical interpretation is
the same as in the case of linear acceleration: The Lorentz-Fitzgerald
contracted arclength between our rotating objects is always $l$,
hence the actual arclength between them must be larger by the 
Lorentz $\gamma$-factor. Again, it is important to recognize that $O_1$ and
$O_2$ could be any two points in a rotating measuring device. 

As in the case of linear acceleration, we now attempt an alternative
description that is also based on the hypothesis of locality and
replace the infinite sequence of momentarily comoving inertial frames by 
one continuously moving frame, for example, the geodesic coordinate
system around the worldline of one of the rotating observers.

Consider a rotating observer in the $(ct,x,y,z)$ coordinate system as
in figure \ref{fig:fig5}. It turns out that the natural orthonormal
tetrad frame of such an observer is given by \cite{mash90}
\begin{subequations}\label{rotframe}
\begin{align}
\lambda^\mu{}_{(0)} &=
\gamma ( 1, -\beta \sin \varphi, \beta \cos\varphi, 0 ) 
\;,\\
\lambda^\mu{}_{(1)} &=
(0,\cos\varphi,\sin\varphi,0)
\;,\\
\lambda^\mu{}_{(2)} &=
\gamma ( \beta, -\sin \varphi, \cos \varphi, 0)
\;,\\
\lambda^\mu{}_{(3)} &=
(0,0,0,1)
\;.
\end{align}
\end{subequations}
where $\varphi$ is the azimuthal angle of the observer such that
$\frac{d\varphi}{dt}=\Omega_0$, $\beta=\nfrac{r\Omega_0}{c}$ and
$\gamma$ is the corresponding Lorentz factor. In this case, the
components of the acceleration tensor \eqref{eq:6} turn out to be
$\nfrac{\vec{a}}{c}=(-\beta\gamma^2\Omega_0,0,0)$ corresponding to the
centripetal acceleration and the rotation
$\vec{\Omega} =(0,0,\gamma^2\Omega_0)$ of the spatial frame with
frequency $\gamma^2\Omega_0$ about the nonrotating triad that
represents ideal gyroscope directions \cite{mash90}. To determine the
proper acceleration length in this case, we note that
$I=\frac{\gamma^2\Omega_0^2}{c^2}$ and $I^*=0$. Thus
$\mathcal{L}=\frac{c}{\gamma\Omega_0}$, where
$\gamma\Omega_0=\frac{d\varphi}{d\tau}$ is the proper rotation
frequency of the observer.

Let us now construct a geodesic coordinate system based on the tetrad
frame \eqref{rotframe} for observer $O_1$, i.e.\ we set
$\varphi=\varphi_1=\Omega_0 t$ in \eqref{rotframe}. In equation
\eqref{eq:7}, the worldline $\bar{x}^\mu(\tau)$ of $O_1$ is therefore
given in $(ct,x,y,z)$ coordinates by
$O_1:(ct,r\cos\varphi_1,r\sin\varphi_1,0)$, where $t=\gamma\tau$ and
$\varphi_1 =\gamma\Omega_0\tau$. Hence equation \eqref{eq:7} implies
that the rotating geodesic coordinate system $(cT,X,Y,Z)$ is related to the
original inertial coordinates $(ct,x,y,z)$ by (compare figure~\ref{fig:fig5})  
\begin{gather}
ct =  \gamma (c T + \beta Y),\quad
x=(X+r) \cos (\gamma \Omega_0 T) - \gamma Y \sin (\gamma \Omega_0
T),\notag\\
y=\gamma Y \cos(\gamma\Omega_0 T)+ (X+r) \sin(\gamma \Omega_0 T), \quad 
Z=z\;.\label{eq:eq42}
\end{gather}

\begin{figure}
\begin{center}\input{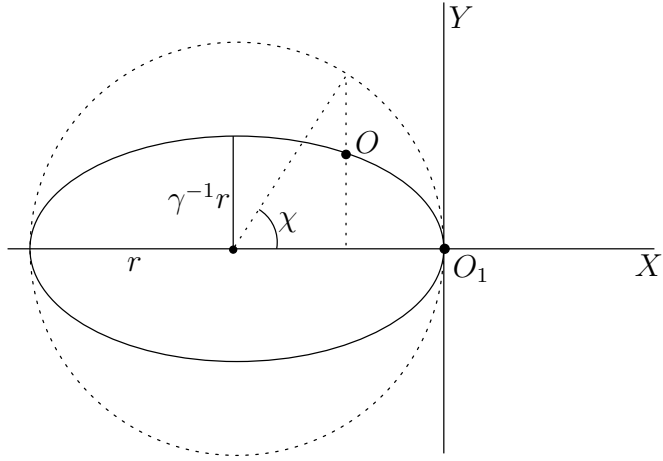}\end{center}
\caption{The observers $O_1$ and $O$ are depicted here from the
standpoint of the geodesic coordinate system established around the
worldline of $O_1$. The ellipse is given by equation \eqref{eq:eq45}
and $O$ would range from $O_1$ at $\chi=0$ up to $O_2$ at
$\chi=\Delta$, where $\Delta - \beta^2\sin\Delta=\Phi$. The length of
the elliptical arc from $O_1$ to $O_2$ is given by $D$ in equation
\eqref{eq:eq47}. This is naturally related to elliptic integrals; that
is, $D=r[E(\frac{\pi}{2},\beta)-E(\frac{\pi}{2}-\Delta,\beta)]$, where
$E(\varphi,k)=\int\limits_0^\varphi \sqrt{1-k^2\sin^2\alpha}\,d\alpha$
is the elliptic integral of the second kind.}
\label{fig:fignew}
\end{figure}

Consider now an observer $O:(ct,r\cos\varphi,r\sin\varphi,0)$ on the
arc between $O_1$ and $O_2$ at a given time $t$ with
$\varphi=\Omega_0t +\phi$ such that the fixed angle $\phi$ could range
from $\phi=0$ at $O_1$ to $\phi=\Phi$ at $O_2$. It follows from the
coordinate transformation \eqref{eq:eq42} that in the geodesic
coordinate system $O:(cT,X,Y,0)$, where 
\begin{equation}
X+r = r\cos\chi\;, \quad Y = \gamma^{-1} r \sin\chi \;. 
\end{equation}
Here $\chi$ is an angle defined by $\chi=\varphi-\gamma\Omega_0T$;
therefore, using $\varphi=\Omega_0t+\phi$ and $t=\gamma T+\gamma\beta
\nfrac{Y}{c}$ we find  
\begin{equation}
\chi-\beta^2\sin\chi=\phi\;. \label{eq:eq44}
\end{equation}
It follows that in the geodesic coordinate system, $O$ lies on an
ellipse 
\begin{equation}
\frac{(X+r)^2}{r^2} + \frac{Y^2}{r^2(1-\beta^2)} = 1 \label{eq:eq45}
\end{equation}
with semimajor axis $r$, semiminor axis $\gamma^{-1}r$ and
eccentricity $\beta=\nfrac{r\Omega_0}{c}$ as depicted in figure
\ref{fig:fignew}. This figure should be compared and contrasted with
figure \ref{fig:fig5}. The ellipse can be thought of as the circle of
radius $r$ that is Lorentz-Fitzgerald contracted along the direction
of motion (i.e.\ the $Y$-axis). The angle $\chi$ is similar to the
eccentric anomaly in Keplerian motion and ranges from $\chi=0$ at
$O_1$ to $\chi=\Delta$ at $O_2$, i.e.
\begin{equation}
\Delta-\beta^2\sin\Delta=\Phi\label{eq:eq46}
\end{equation}
by equation \eqref{eq:eq44}. It is interesting to note that equation
\eqref{eq:eq44} is similar to the Kepler equation for elliptical
motion in Newtonian gravity, except that in the Kepler equation the
eccentricity $\beta$ takes the place of $\beta^2$ in
\eqref{eq:eq44}. Moreover, for a given angle $\phi$, there is a unique
angle $\chi$ for $0\le \beta^2<1$. 

In the rotating geodesic coordinate system established around $O_1$,
the distance from $O_1$ to $O_2$ along the elliptical arc is $D$, 
\begin{equation}
D = r \int\limits_0^\Delta \sqrt{1-\beta^2\cos^2\chi}\,d\chi\;,\label{eq:eq47}
\end{equation}
which is in general different from $l'=\gamma r\Phi$. For instance,
for a fixed $\Phi$, $l'\to \infty$ as $\beta\to 1$, while $D\to
r(1-\cos\Delta)$ in this limit so that $\nfrac{D}{l'} \to 0$. Moreover,
$D$ is a monotonically increasing function of $\Phi$ for fixed
$\beta$. 

On the other hand, let us fix $\Phi$ at $\pi$ and note that when
$\Phi=\pi$, $\Delta=\pi$ as well from equation \eqref{eq:eq46}; then,
the half circumference of the ellipse is given by 
\begin{equation}
D=\pi r \left[ 1- \left(\frac{1}{2}\right)^2 \beta^2 - \left(
\frac{1\cdot 3}{2 \cdot 4}\right)^2 \frac{\beta^4}{3} - \left(
\frac{1\cdot 3\cdot 5}{2 \cdot 4\cdot 6}\right)^2 \frac{\beta^6}{5} -
\mathcal{O}(\beta^8) \right] \;, \label{eq:48}
\end{equation}
so that as $\beta$ goes from $0 \to 1$, the corresponding $D$
\emph{decreases} 
from $\pi r \to 2r$ and $\nfrac{D}{l'}$ goes from $1\to 0$. To
understand this variation intuitively, we note that
$\beta\gamma=\nfrac{r}{\mathcal{L}}$. That is, 
\begin{equation}
\frac{l'}{\mathcal{L}} = \beta\Phi
\end{equation}
in the case under consideration here with $0\le \Phi<2\pi$. Thus, when
the circular orbit is much smaller than the acceleration length of the
observer, $\beta\gamma=\nfrac{r}{\mathcal{L}}\ll 1$, expanding
equation \eqref{eq:eq47} in
powers of $\beta^2\ll 1$ we find that 
\begin{equation}
\frac{D}{l'} = 1 - \frac{3}{4}\beta^2 \left( 1 + \frac{\sin2\Phi - 8
\sin\Phi}{6\Phi}\right) + \mathcal{O}(\beta^4) \;,\label{eq:eq50}
\end{equation}
where $\Phi=\frac{l}{r}$. When the radius of the circular orbit is
much smaller than the acceleration length of the observer, $D\approx
l'$; however, the deviation of $\frac{D}{l'}$ from unity cannot be
neglected for $\beta\to 1$. 

If the geodesic coordinate system is established along the worldline
of the observer $O_2$ instead, then the arclength from $O_2$ to $O_1$
in the accelerated system turns out to be $D$ as well due to the
symmetry of the uniformly rotating configuration depicted in figure
\ref{fig:fig5}. 

Considering our results, it is necessary to recognize that there is no
unique answer 
for event distances when the observer is accelerated. We do not have
a theory that gives us the precise distance on the Earth between
Cologne (Germany) and 
Columbia (Missouri), for example, since the Earth rotates. Of course,
$\epsilon=\beta\gamma\Phi$ is typically very small, since it compares $l$ with the
very large acceleration length $\mathcal{L}$. For instance, for
antipodal points along the equator, equation \eqref{eq:eq50} implies
that the difference between $D$ and $l'$ amounts to only a distance of
the order of
$10^{-3}\,\text{cm}$. 

\section{Discussion}
The main purpose of this work has been to demonstrate that within the
confines of classical, i.e.\ nonquantum, physics there exist basic
limitations on length measurement by accelerated observers in
Minkowski spacetime that follow
from the hypothesis of locality. Indeed, realistic accelerated
coordinate systems suffer from limitations that are far more severe
than those imposed by the requirement of the admissibility of such
coordinates. That is, all distances in accelerated systems must in
fact be negligibly small compared to the characteristic acceleration
lengths of the observer.\label{sec:sec6} 

Discussions of the quantum limitations of spacetime measurements are
contained in \cite{mash89} and \cite{salecker}. Difficulties with the
measurement of spatial distance in the general theory of relativity
are treated in \cite{schmidt}.

\end{document}